\documentstyle[aps,twocolumn,prl,tighten,flushrt,psfig]{revtex} 
\input epsf

\def\be{\begin{equation}}
\def\ee{\end{equation}}
\def\bea{\begin{eqnarray}}
\def\eea{\end{eqnarray}}

\def\sec{\,{\rm sec}}

\def\cmm2{{\,\rm cm^{-2}}}
\def\cm2{{\,{\rm cm}^2}}
\def\cmm3{{\,{\rm cm}^{-3}}}
\def\gcmm3{{\,{\rm g\,cm^{-3}}}}

\def\fun#1#2{\lower3.6pt\vbox{\baselineskip0pt\lineskip.9pt
  \ialign{$\mathsurround=0pt#1\hfil##\hfil$\crcr#2\crcr\sim\crcr}}}

\def\'{^{\prime}}

\def\p3m{P$^3$M}

\def\la{\mathrel{\mathpalette\fun <}}
\def\ga{\mathrel{\mathpalette\fun >}}
\def\fun#1#2{\lower3.6pt\vbox{\baselineskip0pt\lineskip.9pt
  \ialign{$\mathsurround=0pt#1\hfil##\hfil$\crcr#2\crcr\sim\crcr}}}

\begin{document}
\twocolumn[\hsize\textwidth\columnwidth\hsize\csname @twocolumnfalse\endcsname
\title{Probing Unstable Massive Neutrinos with Current
Cosmic Microwave Background Observations}
\author{Robert E. Lopez,$^{1,2}$ Scott Dodelson,$^2$ Robert J. Scherrer,$^{2,3}$
and Michael S. Turner$^{1,2,4}$}
\address{$^1$Department of Physics
The University of Chicago, Chicago, IL~~60637-1433}
\address{$^2$NASA/Fermilab Astrophysics Center
Fermi National Accelerator Laboratory, Batavia, IL 60510, USA}
\address{$^3$Department of Physics and Department of
Astronomy
Ohio State University, Columbus, OH~~43210}
\address{$^4$Department of Astronomy \& Astrophysics
Enrico Fermi Institute, The University of Chicago, 
Chicago, IL~~60637-1433}
\date{\today}
\maketitle

\begin{abstract}
The pattern of anisotropies in the Cosmic Microwave Background
depends upon the masses and lifetimes of the three neutrino species.
A neutrino species of mass greater than 10 eV with lifetime between
$10^{13}\sec$ and $10^{17}\sec$ leaves a very distinct signature (due
to the integrated Sachs-Wolfe effect): the anisotropies at large angles are
predicted to be comparable to those on degree scales. Present data exclude such a
possibility and hence this region of parameter space. 
For $m_\nu \simeq 30$ eV, $\tau \simeq
10^{13}$ sec, we find an interesting possibility: the Integrated Sachs Wolfe peak
produced by the decaying neutrino in low-$\Omega$ models mimics the acoustic peak
expected in an $\Omega = 1$ model.

\end{abstract}
\pacs{Valid PACS appear here.}
]
{\parindent0pt\it Introduction.} The possibility that one or more of the
neutrino species is massive has intrigued cosmologists for over
thirty years\cite{ZELDOVICH}. 
A massive neutrino contributes to the energy density in the
universe. A mass on the order of $50$ eV is sufficient to push the total
density in the universe up to the level expected if the universe is 
flat\cite{SZALAY}.
Masses much larger than this are ruled out cosmologically, based on
measures of the age of the universe. A caveat to this argument is
that an unstable neutrino can decay early enough so that its decay
products redshift away and no longer contribute too much energy to the
universe. This caveat relies on the fact that the energy density
of relativistic particles (such as the decay products) drops faster
than that of non-relativistic particles (such as massive neutrinos).

The caveat that neutrinos with mass much greater than $50$ eV are allowed
cosmologically if they decay fast enough is important, for present 
laboratory limits for the $\mu$ ($\tau$) neutrino are $170$ keV ($24$
MeV)\cite{EXPERIMENT}. The theoretically-predicted
decay modes and lifetime of a 
massive neutrino are very model dependent\cite{NUREVIEW},
with shorter lifetimes typically
arising in familon models wherein 
the decay products are a Majoron and a lighter 
neutrino.  Here we consider primarily such decay modes:  $\nu_H \rightarrow
\nu_L + \phi$, where $\phi$ is a Majoron and $\nu_H$, $\nu_L$ are
any of $\nu_\tau$, $\nu_\mu$, $\nu_e$.  Although our limits
will also be applicable\footnote{Our limits also apply to 3-body decays
as long as the decay products are relativistic.} to photon-producing decays:
$\nu_H \rightarrow \nu_L + \gamma$, there are already other restrictive
limits on such decays.  For the decay $\nu_\tau \rightarrow
\nu_\mu + \phi$, a plausible
lower limit
can be placed on the lifetime from upper limits
on the branching ratio of $\tau \rightarrow \mu + \phi$\cite{LIFE}
\be
\label{lifetime}
\tau > \biggl({0.67~{\rm MeV} \over m_{\nu_\tau}}\biggr)^3 {\rm sec}
\ee

In this {\it Letter}, we argue that a large part of the mass/lifetime
parameter space is strongly disfavored by recent observations of 
anisotropies in the Cosmic Microwave Background (CMB). Dozens of
observations\cite{peakexists} have confirmed that (i) anisotropies exist in the CMB and
(ii) the level of anisotropy is higher on degree scales than on the large
angular scales probed by the COBE\cite{COBE} satellite. Any viable cosmological
model must account for these observations. We show here that a cosmological
model with a decaying
neutrino with mass $m \ga 10$ eV and lifetime in the range $10^{13} {\rm sec}
\la \tau \la 10^{17} {\rm sec}$ fails to produce substantial power on degree
scales and therefore fails to reproduce the current observations. 

Our argument
assumes an otherwise standard Cold Dark Matter (sCDM) model, wherein the total density
total is equal to the critical density ($\Omega = 1$). This density consists of
the energy associated with one species of neutrino with mass $m$ and lifetime
$\tau$ and its relativistic decay products; ordinary baryons with an abundance suggested
by recent measurements\cite{TYTLER} of deuterium $(\Omega_B h^2 = 0.02)$; and
the remainder in the form of cold dark matter. For most of our discussion,
the Hubble constant is fixed at $H_0 = 50\,{\rm km\,sec^{-1}\,Mpc^{-1}}$. Finally
the CMB anisotropy spectrum depends on the primordial spectrum, which we
assume to be Harrison-Zel'dovich, with spectral index $n=1$. We also assume
the decay products of the massive neutrino are sterile, i.e. no photons are produced.
This model serves as a useful framework within which to make our argument. 
After presenting the
argument and the subsequent limits on $m,\tau$, we discuss the changes
in these limits if the true parameters differ from those in our canonical model.

{\parindent0pt\it Integrated Sachs-Wolfe Effect.}

In most theories of structure formation, anisotropies in the CMB today reflect
conditions in the universe at the time of last scattering, when electrons and protons
combined to form neutral hydrogen. After that time (when the universe was of order
$10^{13} {\rm sec}$ old corresponding to a redshift $z \simeq 1100$) photons
travelled freely through the universe. In particular, on angular scales smaller than
a degree, the anisotropies reflect the fact that at last scattering, the combined
electron/ baryon/photon fluid was undergoing acoustic oscillations\cite{HS}. In
Fourier space, the temperature peaked at wavenumber $k_p \sim \pi/r_s = 0.024$
Mpc$^{-1}$ where $r_s$ is the sound horizon at last scattering. If we expand the
anisotropy spectrum today in terms of Legendre polynomials:
\be
C(\theta)
\equiv
\langle
        \Delta(\hat\gamma) \Delta(\hat\gamma') 
\rangle_{\hat\gamma\cdot\hat\gamma'=\cos\theta}
=
\sum_l {(2l+1)\over 4\pi}  C_l P_l(\cos \theta),
\ee
then the power $l(l+1)C_l$ peaks at $l_p \simeq k_p\eta_0 = 200$ where 
$\eta_0\simeq 2H_0^-1 $ is
the conformal time today. In sCDM, the ratio of the power at $l_p$ (on degree scales)
to that at $l\simeq 10$ (COBE scales) is of order $5:1$, and indeed this ratio is
consistent with present data.

While the anisotropy spectrum on small angular scales does indeed derive from
conditions at last scattering, perturbations on larger angular scales enter
the horizon only after this epoch and they are therefore sensitive to conditions
at late times. In sCDM, the gravitational potential due to the cold dark
matter is constant. Coupled with the assumption of a Harrison-Zel'dovich spectrum,
the constant gravitational potentials lead to flat power on large scales. 
Variants of sCDM sometimes predict deviations from this canonical prediction;
in particular, unless the universe is dominated by non-relativistic particles
after last scattering, the gravitational potentials will not in general be constant.
An example of this is an open universe, where curvature begins to dominate at late
times.  If the gravitational potential $\Phi$ is not constant at late times, then
the resulting contribution to anisotropy\cite{HS} on a scale $l$ is
\be
									 \label{eq:ISW}
\Delta_l = -2 \int_{\eta_*}^{\eta_0} d\eta {d\Phi\over d\eta} j_l[k(\eta_0 - \eta)]
\ee
where $\eta_*$ is the epoch of last scattering. The impact of a varying gravitational
potential as described in Eq. \ref{eq:ISW} is called the Integrated Sachs-Wolfe (ISW)
effect. One important feature of Eq.  \ref{eq:ISW} is the factor of two in front;
this is to be compared with the factor of $1/3$ in front of the ordinary Sachs-Wolfe
effect. Even a small decay in the potential can have dramatic implications for the
anisotropies.

If $\Phi$ starts to vary at conformal time $\eta_v$, the scales most affected are
those just entering the horizon at that time, with wavenumbers $k \sim
\pi/\eta_v$. The spherical Bessel function in Eq. \ref{eq:ISW} tells us that these
will be projected onto angular scales of order $l\sim \pi (\eta_0/\eta_v-1)$. It will
be useful to rewrite this order of magnitude estimate in terms of cosmological time
$t$ (instead of conformal time $\eta$). Since $\eta\propto t^{1/3}$ when CDM is the
dominant component, the ISW effect peaks at
\be
\label{eq:EST}
l \sim \pi \left( \big( {4\times 10^{17} {\rm sec} \over t_v} \big)^{1/3} - 1\right)
.\ee Equation \ref{eq:EST} tells us that any disturbance to a flat, non-relativistic
matter dominated universe at times $10^{13} \la t \la 10^{17} {\rm sec}$ will show up
in an enhancement in the anisotropy spectrum leftward of the peak at $l_p \sim 200$.
As we now show, models in which a massive neutrino is unstable with a lifetime in the
above range produce just such a disturbance.

{\parindent0pt\it Anisotropies in Massive Neutrino Models.}

The decay products of a massive unstable neutrino are typically very light, so
decays turn non-relativistic energy into relativistic energy. This leads to a decay
of the gravitational potential: the relativistic particles do not clump as easily as
their massive parents. The decay in the gravitational potential, as we have seen,
leads to an ISW effect at $l$ given by Eq.~\ref{eq:EST} with $t_v$ now replaced by
the massive neutrino lifetime $\tau$. While Eq. \ref{eq:EST} gives the location of
the ISW enhancement, the amplitude of the enhancement is proportional to the ratio of
energy in relativistic decay products to that in CDM. Figure 1 shows an example for a
$10$ eV decaying neutrino with lifetime $\tau=10^{15}$ sec.  There are several
features of note here. First, the energy in decay products peaks at $a \sim
0.02$ which, since $a\propto t^{2/3}$, does indeed correspond to $t_v \simeq
\tau$. Second, decay radiation peaks at only $10 - 20\%$ of the density of
CDM. However, due to the large coefficient in front of the ISW effect, even this
small contaminant of radiation produces a large change in the CMB spectrum.

\begin{figure}[bthp]
\centerline{\psfig{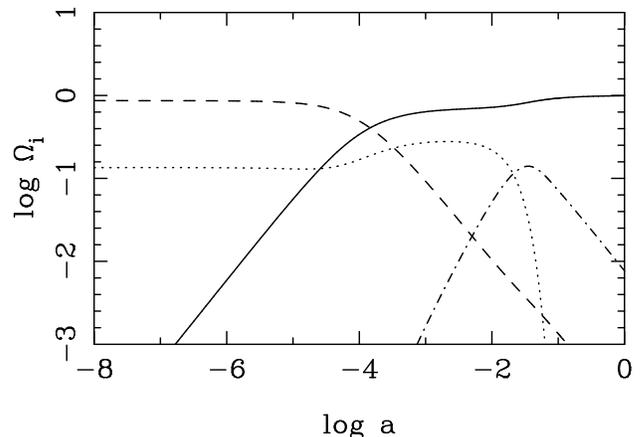}}
\caption[caption]{The densities of the components of the universe as a function of scale
factor for a neutrino with $m=10$ eV and $\tau=10^{15}$ sec. Densities are with respect to the the total density and the 
scale factor is
set to one today. Solid line is CDM which dominates at late times; long dashed line
is radiation in the form of photons and two light neutrino species; short dashed line
is the density of relativistic decay products of the massive neutrino; and the dotted
line is the massive neutrino.}
\label{fig1}
\end{figure}

Figure 2 shows the power spectra of five different models, normalized by the peak at
$l\sim 200$.  The canonical curve is sCDM, for which the 
peak $(l = 200)$/plateau $(l=10)$
ratio  is $5.6/1$. The other curves are decaying
neutrino models, each with a mass of $10$ eV.  As the neutrino lifetime gets longer,
the peak due to the ISW effect moves out to smaller $l$, in accord with
Eq.~\ref{eq:EST}. Note that the ISW peak is quite substantial even though a mass this
small produces relatively little energy in relativistic decay products.

\begin{figure}[bthp]
\centerline{\psfig{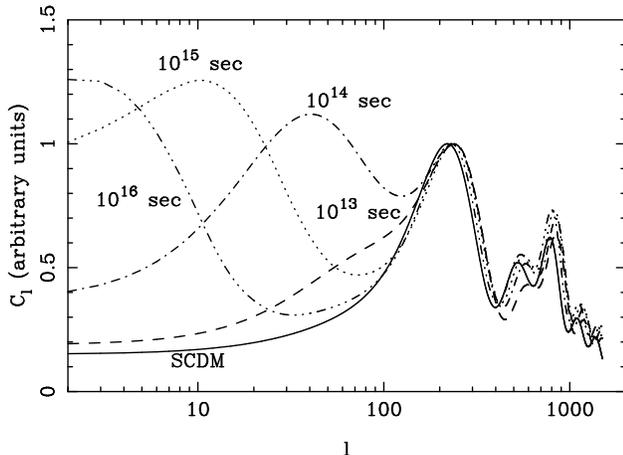}}
\caption[caption]{CMB Anisotropy Spectra for Decaying Neutrino
models. Solid line is standard CDM with three species of
massless neutrinos. Curves are normalized at the first
peak at $l\sim 200$. All curves are for a neutrino
with $m=10$ eV.} 
\label{fig2}
\end{figure}

Larger mass neutrinos produce larger ISW effects. Hence
$m \ga 10$ eV is ruled out if the lifetime is such
that the ISW effect is to the left of the peak. Figure 3 shows
a contour plot of the peak/plateau ratio in the $m,\tau$ plane.
Also shown is the curve corresponding to $\Omega = 1$, and the
bound on $\tau$ from equation (\ref{lifetime}).
A rough, very conservative cut would be to disallow the region
in which the ratio is less than one. As seen from Figure 3, this
cut excludes a large region of parameter space that would
otherwise be allowed.

A number of other constraints can be
placed on such long-lived neutrinos.  If such neutrinos
or their decay products dominate the density, they can yield
an age for the universe which is below bounds from globular
cluster ages \cite{ST}.  In Figure 3, these limits lie
near the $\Omega=1$ curve.  More restrictive constraints from structure formation
are also possible \cite{ST} but these are considerably
more model-dependent than our very simple result. 

\begin{figure}[bthp]
\centerline{\psfig{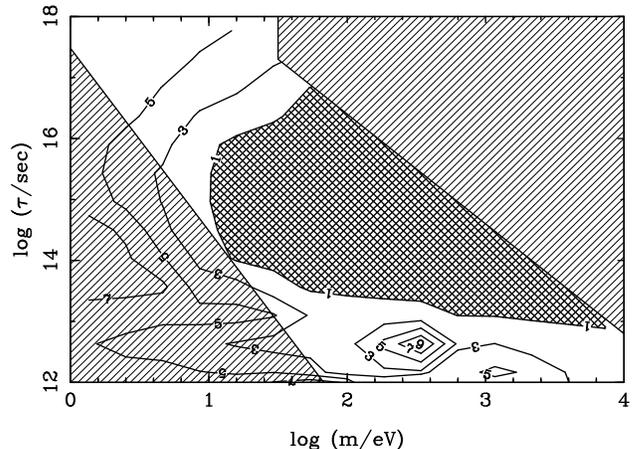}}
\caption[caption]{Contours of the 
peak $(l = 200)$/plateau $(l=10)$ ratio
for CMB spectra. sCDM predicts a ratio on the order of five;
data certainly excludes values less than one. The upper right
region of parameter space produces $\Omega
> 1$.  The lower left hatched region is the bound on
$\nu_\tau \rightarrow \nu_\mu + \phi$
from equation
(\ref{lifetime}).}
\label{fig3}
\end{figure}

{\parindent0pt\it Model Dependence of the Constraint.} The contours drawn in Figure 3
depend on the underlying cosmological model and associated set of parameters. How
does the constraint hold up as these vary? Certainly, all of this work is predicated
on the assumption that the primordial perturbations are adiabatic. This assumption at
present seems well-motivated both theoretically and observationally\cite{LIDDLE}.  
Raising the Hubble constant to a value more consistent with present data lowers the
peak. A higher $\Omega_B$ would help here, but our value of $0.08$ is about as high
as possible, given constraints from Big Bang Nucleosynthesis\cite{TYTLER}.  We have
also neglected reionization, but this too lowers the peak/plateau ratio even
further. We have neglected tensor modes, but these too work in the wrong direction,
falling off at $l \ga 100$.  Our model assumes that the decay products are sterile.
However, our results also apply to photon-producing decays which occur after last
scattering; such decays would also lead to a large ISW effect.  But in the parameter
range of interest, the photons produced in such decays would seriously distort the
well-measured\cite{YDIST} thermal spectrum of the CMB or produce an (unobserved)
diffuse photon background\cite{KT}.

One parameter which could serve to produce a rise on small scales
is the spectral index of the
primordial perturbations, $n$, which we have set to one.  Most inflationary
models\cite{INFLATION} predict values slightly smaller than one, which would again
reduce even further the peak/plateau ratio. However, there are some models
(e.g. hybrid inflation\cite{HYBRID} and supernatural inflation
\cite{SUPERNAT})
which predict ``blue'' spectra ($n>1$). Even if we allow $n>1$, though,
the vast majority of the excluded
$m,\tau$ region in Figure 3 will remain excluded. To see why, consider Figure 4,
which shows the spectrum for a set of excluded $m,\tau$ values. 
The spectrum can be modified to go through the data at $l\sim 200$ if $n=2.15$.
Such an extreme value of $n$, however, is strongly disfavored at both
higher and lower values of $l$.

\begin{figure}[bthp]
\centerline{\psfig{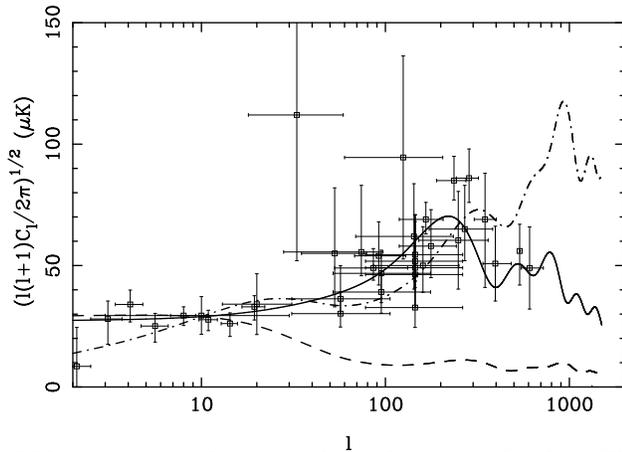}}
\caption[caption]{Impact of varying the scalar spectral index. The solid line is the sCDM
prediction. The dashed line is for a neutrino with $m=30$ eV and $\tau=10^{15}$ sec,
while the dashed-dotted line is the same decaying neutrino model, but with a scalar
spectral index $n=2.15$. This value was chosen to make the CMB exhibit a rise at
$l\sim (100-200)$, as indicated by observations, shown here by the points. Note the
poor fit of the tilted model at low and high values of $l$.}
\label{fig4}
\end{figure}

Finally, we have 
assumed a flat universe today ($\Omega=1$). An open universe, however, would only
exacerbate this effect since the peak/plateau ratio is already relatively small in
open models. This is true for two reasons. First, curvature-induced gravitational
potential decays cause extra power at $l\sim 10$. Second, lowering $\Omega$ shifts
the first acoustic peak to the right, lowering the anisotropy at $l\sim200$. 
There is one interesting caveat to this argument. In an open universe,
the ISW effect due to the decay products also shifts to larger $l$. For small
enough lifetimes, the ISW peak could be shifted all the way to $l\sim 200$,
mimicking the first acoustic peak in a flat universe!
For example, a model with $m=30$ eV, $\tau = 10^{13}$ sec and $\Omega =
0.3$ gives an ISW peak near $l=200$ and a peak/plateau ratio $\sim 6$, in rough
agreement with current observations.

{\parindent0pt\it Conclusions.} An unstable neutrino with mass greater than $10$
eV and lifetime  $10^{13} \la \tau \la 10^{17}$ sec is ruled out by
current CMB observations. These limits are quite robust to changes
in the underlying cosmological model.
Moreover, they apply generally to any massive particle species (e.g.
a neutralino or gravitino)
that contributes significantly to the energy density
and decays in the post-decoupling time frame.
They also rule out the possibility of a radiation dominated
Universe, proposed in 1984 to solve the $\Omega$ problem \cite{RADDOM}.
Moving beyond current observations, 
precision measures of the CMB spectrum
out to $l \sim 2000$, as expected from the MAP and PLANCK satellites\cite{SAT}, are
capable of probing the mass/lifetime plane for all $m \ga 1$ eV\cite{MASTERPIECE}.

\vskip 0.2in
\noindent We thank U. Seljak and M. Zaldariagga for the use of 
CMBFAST\cite{CMBFAST}, which we amended to include decaying
neutrinos. 
This work is supported by the DOE and by NASA Grant NAG 5-7092.

\end{document}